\documentclass[10pt]{article}

\def\bvec#1{{\rm\bf #1}}
\def\meanflow{u_0}

\begin{document}
\title{\large\bf New analytic solutions 
of the non-relativistic hydrodynamical equations}
\author{\normalsize P. Csizmadia, T. Cs{\"o}rg{\H o} and B. Luk{\'a}cs \\[0.5ex]
\small{MTA KFKI RMKI, H - 1525 Budapest 114, POB 49, Hungary}}
\date{}
\maketitle

\begin{abstract}
New solutions are found for the non-relativistic hydrodynamical equations.
These solutions describe expanding matter with a Gaussian density profile.
In the simplest case, thermal equilibrium is maintained {\it without any
interaction}, the energy is conserved, and the process is isentropic.
More general solutions are also obtained that describe explosions driven
by heat production, or contraction of the matter caused by energy loss.
\end{abstract}

{\it Introduction.} 
The equations of hydrodynamics correspond to local conservation
of some charges as well as energy and momentum. 
The equations are scale-invariant, hence can be applied 
to phenomenological description of physical
phenomena from collisions of heavy nuclei to collisions of galaxies.
Recently, a lot of experimental and theoretical effort went into 
the exploration of hydrodynamical behaviour of strongly interacting
hadronic matter in non-relativistic as well as in relativistic
heavy ion collisions, see for example refs. \cite{csernai}-\cite{qm96}.
Due to the non-linear nature of the equations of hydrodynamics,
exact solutions of these equations are rarely found.
In ref. \cite{jozso}
an exact solution of hydrodynamics of expanding fireballs was found 20 years
ago.
The purpose of this Letter is to present and analyze
a new, exact solution of the non-relativistic hydrodynamical equations,
with a generalization to heat production or loss (e.g. due to radiation).
We hope that the results presented herewith may be utilized to
access analytically the time-evolution of the hydrodynamically
behaving strongly interacting matter as probed by non-relativistic
heavy ion collisions \cite{nr,nrt}. The results are,
however, rather general in nature and they
can be applied to any physical phenomena where the non-relativistic
hydrodynamical description is valid.

{\it Adiabatic expansion.}
Consider a hydrodynamical system described by the {\it continuity equation},
the {\it Euler equation} and the local {\it energy conservation}:
\begin{eqnarray}
{\partial n\over\partial t} + \bvec\nabla (\bvec v n) &=& 0,
  \label{continuity-eq}\\
\left({\partial\over\partial t} + \bvec v\bvec\nabla\right) \bvec v
 &=& -{\bvec\nabla P\over m n},\label{Euler-eq}\\
{\partial\varepsilon\over\partial t} + \bvec\nabla(\bvec v \varepsilon)
 &=& - P\bvec\nabla\bvec v.\label{energy-eq}
\end{eqnarray}
We close this set of equations with the help of an ideal gas equation of
state,
\begin{eqnarray}
\varepsilon(\bvec r,\, t) &=& {3\over 2} P(\bvec r,\, t),\label{energy-density}\\
P(\bvec r,\, t) &=& n(\bvec r,\, t) T(t).\label{EOS}
\end{eqnarray}
These equations are solved by 
\begin{eqnarray}
n(\bvec r,\, t)
&=&  {N\over (2\pi R^2(t))^{3/2}}\, \exp\left(-{r^2\over 2R^2(t)}\right),
\label{particle-density}\\
T(t) & = & {T_0\over\varphi(t)},\qquad
R^2(t)\ \ =\ \ R_0^2\varphi(t),\qquad
\bvec v(\bvec r,\, t)\ \ =\ \ {\dot\varphi(t)\over 2\varphi(t)}\bvec r,
\label{new-parameters}\\
\varphi(t) &=& \left(1 + {t-t_0\over\tau}\right)^2
              + {T_0\over T_G} \left({t-t_0\over\tau}\right)^2,\qquad
T_G\ \ =\ \ m u_0^2,
\label{phi-in-new-parameters}
\end{eqnarray}
where $u_0=R_0/\tau$ is the characteristic expansion rate.
This is a new solution of the non-relativistic hydrodynamical
equations \cite{csernai}. The flow field has the same distribution as in the
solutions found by Zim\'anyi, Bondorf and Garpman 20 years ago
\cite{jozso}, however the temperature and the density profiles
are different.

Now consider the phase-space distribution function $f$
belonging to this hydrodynamical solution.
This distribution has a locally Maxwell-Boltzmann (MB) shape
everywhere, all the time:
\begin{equation}
f(\bvec r,\, \bvec p,\, t)
\, =\, C\exp\left(-{r^2\over 2R^2(t)}
  - {(\bvec p - m\bvec v(\bvec r,\, t))^2\over 2m T(t)}\right),
\quad C\, =\, {N \over (4\pi^2 m T_0 R_0^2)^{3/2}}.
  \label{free-streaming-solution}
\end{equation}

Suppose that the distribution is created at $t=t_0$ by an instantaneous source.
It solves the following collisionless Boltzmann equation for $t>0$:
\begin{equation}
\left({\partial\over\partial t} + \bvec v\bvec\nabla\right) f
\, =\, {\cal S}(\bvec r,\, \bvec p,\, t)
\, =\, C \exp\left(- {r^2\over 2R_0^2}
- {\left(\bvec p-{m\over\tau}\bvec r\right)^2\over 2mT_0}\right) \delta(t-t_0).
\label{emission-function}
\end{equation}
Since the system is considered after $t_0$, 
the normalization coefficient $C$ 
is constant because $T(t) R^2(t)=T_0 R_0^2=const.$
Solutions to the collisionless Boltzmann equation
are not uncommon, see e.g.~\cite{bsol},
however the cited solutions generate deviations from the MB shape
of the local momentum distributions, hence
they are not solutions of the hydrodynamical equations.

In contrast, assuming here a special initial condition,
spherical symmetry and a radial initial flow pattern,
i.e. eq.~(\ref{emission-function}),
we find a solution of the collisionless Boltzmann equation
that maintains a MB form everywhere for all times.
In this sense, our solution is really a solution of
hydrodynamics. Physically this corresponds to a gas
that first had been thermalized in equilibrium.
Later, the gas was frozen out in an idealized break-up at $t = t_0$.
Our local phase-space distribution function $f(\bvec r,\bvec p,t)$
maintains its locally thermalized shape without any collisions,
due to the special choice of the initial conditions and
the invariance of the Gaussian shape under convolution.

We find a non-vanishing pressure maintained without collisions;
the interpretation of this result is that any wall or bubble inserted into 
this expanding Knudsen gas would feel a pressure that arises due to the random, 
locally disordered motion of the free-streaming particles in any
part of space. Such a pressure arising from a collisionless 
gas of photons is well-known in cosmology as a source of gravity,
see ref. \cite{Ehlers}.

The {\it entropy density} can be evaluated from kinetic theory,
using (\ref{free-streaming-solution}):
\begin{eqnarray}
s(\bvec r,\, t) &=&\left({r^2\over 2R^2(t)}
 - \ln N + {3\over 2}\ln(4\pi^2 m T(t) R^2(t))
 + {5\over 2}\right) n(\bvec r,\, t).\label{entropy-density}
\end{eqnarray}

By introducing the ``effective volume'' $V_G=(2\pi)^{3/2}R_0^3$
and integrating the entropy density over space, we find that the
total entropy is 
\begin{equation}
S(t) \ =\ S_{\rm ideal}+{3\over 2}N\ =\ const.
\end{equation}
As it was shown in \cite{modify-entropy}, one can modify the
thermodynamical definition of the entropy by adding terms linear
in extensives, without changing the thermodynamics.
Above, we got the interesting result that the entropy is almost the same as
$S_{\rm ideal}$, the entropy of an ideal gas 
at temperature $T_0$ in volume $V_G$
--- the difference is the extensive quantity ${3\over 2}N$,
hence the thermodynamics of the system considered is
the same as that of an ideal gas.
Note that the total entropy is independent of time.
It is easy to verify from eqs.
(\ref{continuity-eq}, \ref{new-parameters}, \ref{entropy-density})
that the process is {\it locally isentropic}:
\begin{equation}
{\partial s\over\partial t}+\bvec\nabla (\bvec v s)\ =\ 0.
\label{isentropic-eq}
\end{equation}

Without rescattering and other final state interactions 
after particle emission, the {\it momentum spectrum} is independent of time:
\begin{equation}
N_1(\bvec p)\, =\, N_1(\bvec p,\, t)
\, =\, {N\over (2\pi m T_*)^{3/2}}\exp\left(-{p^2\over 2 m T_*}\right),
\qquad T_*\, =\, T_0+m u_0^2.
\end{equation}

It is worthwhile to evaluate the total energy in local disordered motion
(heat energy, denoted by $E_{\rm heat}$), 
the total energy in ordered motion (flow 
energy, denoted by $E_{\rm flow}$) and the total energy $E_{\rm tot}$.
One obtains that
\begin{equation}
	E_{\rm tot}\, =\, \frac{3}{2} N T_* = const, \,\quad
	E_{\rm heat}\, =\, \frac{3}{2} N T(t), \,\quad
	E_{\rm flow}\, =\, E_{\rm tot} - E_{\rm heat},
\end{equation}
where the time dependence of the local temperature is given by eqs.
(\ref{new-parameters}-\ref{phi-in-new-parameters}).

The {\it two-particle spectra} $N_2(\bvec p_1,\bvec p_2,
t_1,t_2)$ are also independent of time if $t_1,t_2 > t_0$.
The two particle distribution function can be written in terms of the
emission function $S$ in the following way:
\begin{eqnarray}
N_2(\bvec p_1,\, \bvec p_2,\, t_1,\, t_2)
&=&\intop_{-\infty}^{t_1}\!\! dt_1'\! \int\! d^3r_1\!\!
   \intop_{-\infty}^{t_2}\!\! dt_2'\! \int\! d^3r_2\,
   (S(\bvec r_1,\, \bvec p_1,\, t_1') S(r_2,\, \bvec p_2,\, t_2')\nonumber\\
&\pm& S(\bvec r_1,\, \bvec K/2,\, t_1') S(\bvec r_2,\, \bvec K/2,\, t_2')
      \cos \left[\bvec q(\bvec r_1-\bvec r_2)\right],\nonumber
\end{eqnarray}
where $\bvec K=(\bvec p_1+\bvec p_2)/2$, $\bvec q=\bvec p_1-\bvec p_2$
and the ``+'' sign refers to bosons, ``-''to fermions,
if the final state Coulomb and other interactions are negligible.
$S$ vanishes for $t_{1,2}'>t_0$ so the the time integration
is independent of the upper bound $t_{1,2}$ as long as $t_{1,2}>t_0$.
Therefore $N_2$ is independent of $t_1$ and $t_2$.

Because both the single and the two-particle spectra are independent of time,
the two-particle {\it correlation function}
must also be {\it independent of time}:
\begin{equation}
C_2(\bvec p_1,\,\bvec p_2)
\ =\ {N_2(\bvec p_1,\, \bvec p_2)\over N_1(\bvec p_1) N_2(\bvec p_2)}
\ =\ 1 + e^{-R_*^2\bvec q^2}.
\end{equation}
So the radius parameter of the correlation function,
$R_*$ is time independent.
We can calculate it at $t_0$, the time of particle production.
Let $u_0=R_0/\tau$. The result
\begin{equation}
{1\over R_*^2}\ =\ {1\over R_T^2} + {1\over R_0^2}
\ =\ {1\over R_0^2}\left(1+{m\meanflow^2\over T_0}\right)
\end{equation}
is a generalization of the result in ref. \cite{nr} for arbitrary values
of $\meanflow$.

{\it More general solutions.}
The above presented solution of the non-relativistic hydrodynamical equations
can be generalized in many ways.
For instance, a straightforward way would be to introduce an
inhomogeneous temperature profile \cite{simple-analytic-gradT}.
Such a temperature profile is present in a known analytic solution,
see ref.~\cite{jozso}.
Here we will investigate the possibility in our model to describe the
effects of some local heat production or heat loss.
A simple model is introduced to mimic the effects of heat
production by chemical or nuclear reactions, or
the cooling of the system by radiation that decreases
the local energy density. 
The microscopic details of these processes are
not sought for and the sources of heat production,
or the radiated energies are not part of the system under consideration,
so heat production of radiation changes the `total' energy.

Let us introduce an additional term into the energy balance 
equation (\ref{energy-eq}).
\begin{equation}
{\partial\varepsilon\over\partial t}+\bvec\nabla(\bvec v\varepsilon)
+P\bvec\nabla\bvec v\ =\ {3\over 2} j(t) n(\bvec r,\, t)T(t).
\label{energy-violating-eq}
\end{equation}
This new term is the simplest possible model of heat production
($j(t)>0$) or energy loss e.g. due to radiation ($j(t)<0$).
The heat loss or heat production is assumed to be proportional
to the local internal energy.

Now we have --- in place of eqs. (\ref{new-parameters}) --- a more general
Gaussian solution of the continuity equation:
\begin{equation}
T(t)\ \ =\ \ T_0 {h(t)\over\varphi(t)},\qquad
R^2(t)\ \ =\ \ R_0^2\varphi(t),\qquad
\bvec v(\bvec r,\, t)\ \ =\ \ {\dot\varphi(t)\over 2\varphi(t)}\bvec r,
\label{nonisentropic-new-parameters}
\end{equation}
where $h$ is so far an unknown function.
This parameterization satisfies the boundary conditions for the
temperature and the radius if $\varphi(t_0)=h(t_0)=1$.

By substituting the expressions of
the energy density (\ref{energy-density}),
the pressure (\ref{EOS}) and
the particle density (\ref{particle-density})
into the modified energy equation (\ref{energy-violating-eq}),
and using the parameterization (\ref{nonisentropic-new-parameters}),
we get the following differential equation for $h$:
\begin{equation}
\dot h(t)\ =\ h(t) j(t).\label{h-diff-eq}\label{dhdt}
\end{equation}
Note that adding source terms to the energy 
balance equation results is a deviation from the isentropic
expansion or contraction (\ref{isentropic-eq}). 
From (\ref{entropy-density}), (\ref{continuity-eq}),
(\ref{nonisentropic-new-parameters}) and (\ref{h-diff-eq}),
the local entropy production is
\begin{equation}
{\partial s\over\partial t} + \bvec\nabla (\bvec v s)\ =\ {3\over 2}j n.
	\label{e:sprod}
\end{equation}

The solution of eq. (\ref{h-diff-eq}) is
\begin{equation}
h(t)\ =\ e^{\intop_{t_0}^t dt' j(t')}. \label{e:hsol}
\end{equation}
The function $\varphi(t)$ can be determined from the Euler equation,
\begin{equation}
\ddot\varphi(t)\varphi(t)-{1\over 2}\dot\varphi^2(t)
\ =\ {2T_0\over m R_0^2}h(t).
\label{violating-phi-diff-eq}
\label{e:ht}
\end{equation}
Supposing $j(t)=const\neq 0$, equation (\ref{violating-phi-diff-eq})
has two exponential solutions:
\begin{equation}
\varphi(t)\ =\ e^{{1\over 2}j\cdot(t-t_0)},\qquad
h(t)\ =\ e^{j\cdot(t-t_0)},\quad
j\ =\ \pm\sqrt{16 T_0\over m R_0^2}.
\end{equation}
In terms of temperature, radius and velocity, the solution reads as:
\begin{equation}
T(t)\ =\ T_0 e^{{1\over 2}j (t-t_0)},\qquad
R^2(t)\ =\ R_0^2 e^{{1\over 2}j (t-t_0)},\qquad
\bvec v(\bvec r,\, t)\ =\ {j\over 4}\bvec r.
\end{equation}

In case of $j>0$, we have an exponential expansion and warming up
driven by an external energy source.
The opposite case is the exponential contraction of the matter, for $j<0$.
It may be caused by the continuous emission of energy by radiation.
The flow field is independent of time in both cases.

{\it Algorithm to generate new solutions}.
One can generate even infinitely many new
analytical solutions to the equations of
non-relativistic 
hydrodynamics with energy producing or radiative processes,
using the following method:

{\it 1)} Assume a functional form for $\varphi(t)$ and choose
$(T_0, m, R_0^2)$ so that
\begin{equation}
\ddot\varphi(t_0)-{1\over 2}\dot\varphi^2(t_0)\,\ =\,\ {2T_0\over m R_0^2},
\qquad\varphi(t_0)\,\ =\,\ 1.
\end{equation}

{\it 2)} Determine the function $h(t)$ from eq.~(\ref{e:ht}).

{\it 3)} Find the energy source function $j(t)$ from eq. (\ref{dhdt}).

This way, a solution of eqs.
(\ref{continuity-eq}-\ref{Euler-eq}, \ref{energy-violating-eq}) is generated.
The only requirement for consistency arises from 
$T(t) > 0$, which results in $h(t) / \varphi(t) > 0$.
The solution is given by eq. (\ref{nonisentropic-new-parameters})
with $\varphi(t)$, $h(t)$ and $j(t)$ generated by steps {\it 1)}-{\it 3)}.

{\it Summary.}
We found a new class of solutions for the non-relativistic hydrodynamical
equations, describing a spherically expanding
Knudsen gas with time dependent but location independent temperature.

Then we incorporated into our formalism the possibility to describe
the effects of the emission or absorption of heat.
In case of the simplest analytic solution, the system expands or
contracts exponentially, while the flow field becomes independent of time
in this special case. 

Finally, we presented an algorithm that generates infinitely
many new analytical solutions of the hydrodynamical equations
with energy sources.

{\it Acknowledgements:} The idea of this paper emerged
in a conversation with M. Pl\"umer at the Columbia University.
T. Cs{\"o}rg{\H o} thanks to L. P. Csernai and J. Zim{\'a}nyi for
stimulating discussions.
This work has been supported by the US-Hungarian Joint Fund,
MAKA 652/1998, by an NWO - OTKA grant N025186 as well
as by an OTKA grant T026435.

\end{document}